\documentclass[runningheads]{llncs}
\usepackage[T1]{fontenc}
\usepackage{booktabs}
\usepackage{amsmath}
\usepackage{makecell}
\usepackage{url}
\usepackage{graphicx}
\usepackage{subcaption}
\usepackage{cite}
\usepackage{graphicx}
\usepackage{color}
\usepackage{soul}

\newcommand{\optlevel}{\textit{optimization\_level} }
\newcommand{\reslevel}{\textit{resilience\_level} }

\begin{document}
\title{Benchmarking the Lights Out Problem on Real Quantum Hardware}
\titlerunning{Benchmarking the Lights Out}
\author{Maksims Dimitrijevs\inst{1}\orcidID{0000-0002-4225-7889} \and
Maria Palchiha\inst{2} %\orcidID{0000-1111-2222-3333}
\and
Abuzer Yakary{\i}lmaz\inst{1}\orcidID{0000-0002-2372-252X}}
\authorrunning{Dimitrijevs, Palchiha, Yakary{\i}lmaz}
\institute{Center for Quantum Computer Science, University of Latvia, Latvia \and
Riga Purvciems Secondary School, Latvia
\email{\{maksims.dimitrijevs,abuzer.yakaryilmaz\}@lu.lv}\\
}
\maketitle              % typeset the header of the contribution
\begin{abstract}
We implement the Lights Out problem on a 2D grid and on M\"obius ladder graphs and evaluate the performance of Grover’s search on real quantum hardware. We use two instances using 9 and 16 qubits, and implement them on publicly available quantum hardware by IBM and IQM. Our experiments show improvements in IBM hardware between the Heron r1 and Heron r2 generations, highlighting progress in IBM hardware during the 2023–2024 period. The Lights Out circuits produced output distributions close to uniform on IQM devices. To diagnose device limitations, we additionally ran a small Grover SAT baseline, finding that IQM Garnet performs more reliably than other tested IQM devices. We also observed that QPUs of the same manufacturing revision can differ significantly in performance (a newer device is not guaranteed to be better), and that calibration has a significant impact on the performance of quantum devices, so the choice of device strongly depends on calibration quality.

%We did not obtain high success probabilities on IQM devices for solving our instances of Lights Out problem, so we did additional benchmarking with Grover's Search implementation with oracle for small case of SAT problem, and concluded that IQM Garnet consistently performs better than IQM Emerald and IQM Sirius.
%\Abu{ChatGPT suggested format to present the results: Our experimental results show that this instance performs well on this quantum device, while it does not yield comparable results on the other device.} 

\end{abstract}
\section{Introduction}

%\Abu{Application-oriented benchmarking}

Publicly available quantum computers have existed for about a decade. Although the number of qubits has increased to over 100, we are still in the NISQ era; that is, currently available hardware is noisy and offers limited qubit connectivity. As a result, researchers have benchmarked various quantum algorithms using noisy simulators as well as real quantum hardware \cite{ref_mills, ref_lubinski, ref_kordzanganeh}.
%Quantum computers are reality, and last years the interest around the possibilities for a quantum computer to solve practical tasks is increasing. Currently we are in NISQ era, meaning, that possibilities for existing quantum computers to solve tasks are limited, mainly because of noise. To measure the progress of development of quantum hardware, different benchmarks are uses. Among benchmarks are: Quantum Volume, largest GHZ state, Q-score, CLOPS, 2Q error (layered). Researchers also investigate small cases of practical algorithms to test the existing hardware (links required).

%\Abu{Publically accessible quantum hardware}

We use publicly accessible quantum hardware. In January 2026, IBM provides 10 minutes of free runtime per month on 3 devices, while IQM gives access to roughly 1 minute of runtime per month on 3 devices. We focus on Grover's Search with oracle implementations for the Lights Out problem to evaluate the performance and practical capacity of these quantum computers. We first check the different versions of the Lights Out problem and design a reasonably small problem instance on a 2D grid and M\"obius Ladder graphs. We design oracles for these problem instances as quantum circuits, and with complete Grover's Search implementation, we have quantum circuits with 9 and 16 qubits, both with similar depths and number of two-qubit operations. These circuits are our primary benchmarks. Additional small Grover's Search circuits are included only as diagnostic baselines to interpret hardware-limited results.

\newcommand{\marrakesh}{\texttt{ibm\_marrakesh}}
\newcommand{\fez}{\texttt{imb\_fez}}
\newcommand{\torino}{\texttt{ibm\_torino}}

%\Abu{Here I defined IBM machines with texttt font sytle. If you like it, you can apply globally.}

We run these instances of Grover's Search on available IBM and IQM hardware -- two Heron r2 devices \marrakesh~and \fez, one Heron r1 device \torino, IQM devices Emerald and Garnet with superconducting transmon qubits on a square lattice, and IQM star topology quantum processor Sirius. For IBM and IQM devices, we try different optimization levels with Qiskit code, and for IBM devices, we try the IBM Composer that is available on the platform. On IQM devices, the Lights Out circuits produced near-uniform outputs; therefore, we decided to run two diagnostic baselines (a small SAT instance on 5 qubits and a two-qubit sanity check) on these devices to check the effects of Grover's Search.
%For IQM devices, we first tested our smallest case of Grover's Search, and as we did not obtain high success probabilities, we decided to test IQM devices on simpler problem - we designed SAT oracle with two variables and two clauses, and resulting Grover's Search implementation used up only 5 qubits.

With our experiments, we concluded that IBM hardware improved over the year period 2023-2024, as newer devices demonstrated higher reliability and better performance. For IQM devices, we observed that their architecture produces more efficient circuits after the transpilation, compared to IBM devices. We also noticed that revision (technology) alone does not guarantee better performance of newer devices, as one Heron r2 device sometimes performed worse than the Heron r1 device, while another Heron r2 device was consistently better than the other two. Moreover, manufacturing quality is not the only contributor to the performance, as we noticed the difference in performance between different calibrations.

%TO DO probably need to rephrase "solvable", "hardest" etc.
We conclude that our Grover's Search instances are suitable for benchmarking the current hardware, as they demonstrate the difference in performance of available quantum computers. Our instance on M\"obius Ladder can almost be solved on available quantum computers, meaning that it can be an interesting benchmark for near-future freely available quantum computers. Interestingly, the instance on a 2D grid has similar depth and total number of two-qubit operations as the instance on M\"obius Ladder, while the difference is in the number of qubits involved, highlighting an interesting case, where performance is not determined solely by gate error rates or by the total number of qubits.

The codes, results of our experiments, and device calibration data files are publicly available at \url{https://github.com/infenrio/lights_out_quantum}.

\section{Grover's Search and Lights Out}

\subsection{Grover's Search}

%\Abu{ChatGPT re-written of the following paragraph: 

Grover’s search provides a quadratic speedup over classical algorithms for unstructured search problems. Many practical problems can be formulated as oracles with binary outputs, so that valid solutions correspond to marked inputs. Implementing such oracles as quantum circuits opens up the possibilities to use Grover’s Search as a scalable benchmarking framework. In this setting, the circuit depth and qubit count naturally scale with the size of the problem, making Grover-based implementations well-suited for evaluating the performance of current quantum hardware \cite{ref_abughanem}.%}
%\Abu{This paragraph seems to need some REF!}

%Grover's Search \st{ a quantum algorithm created by Lov K. Grover, and it } provides quadratic speedup over classical algorithms on unstructured data search (https://dl.acm.org/doi/10.1145/237814.237866). The algorithm can be used to find the solution to some problem, that can be defined as an oracle. We can look at the oracle as a function, that computes the output value for the provided input. If our problem has a binary output, and input data that results in function value 1 is solution to our problem, then implementation of the function with quantum operations can serve as an oracle to the Grover's Search. This gives us scalable framework to implement quantum circuits for practical tasks that will serve as benchmarking for available quantum computers. Grover's Search implementation scales both for number of qubits and depth of the circuit - parameters, that comprehensibly represent the performance of quantum computers.

%\Abu{One smooth transition sentence and then you can give directly the following examples with their references.}

Motivated by this benchmarking potential, researchers have tested quantum computers using Grover’s Search for instances of different computational problems. For example, implementations of Grover's Search for different SAT problems highlighted that small instances of problems are difficult for available devices \cite{ref_vinod, ref_bennakhi}. Different designs of the SAT problem were considered to improve the solvability of the problem with existing quantum hardware \cite{ref_alemany}.

The work in \cite{ref_joshi} provided a detailed observation about how a quantum computer’s coupling map (qubit connectivity graph) and noise influence Grover’s algorithm. The work in \cite{ref_akmal} showed that simpler architectures can yield higher success probabilities than bigger but noisier ones.

While Grover’s quadratic speedup may be lost due to noise, its benchmarking remains interesting, and it is applied to the new problems \cite{ref_jipipob, ref_haverly}.

\subsection{Lights Out}

The Lights Out game appeared as an electronic puzzle game in 1995, released by Tiger Electronics \cite{ref_anderson}. The game field consists of lamps that are arranged in a grid $5 \times 5$. When the game starts, some of the lamps are switched on. When a lamp is clicked, the state of it and its adjacent neighbors is changed (on $\leftrightarrow$ off). The objective is to switch all the lamps off. The game can also have an additional objective, such as using the minimal possible moves.

The Lights Out game has been popular within the Mathematics and Computer Science communities, as it represents an interesting combinatorial puzzle. It was mathematically investigated in \cite{ref_anderson}. In \cite{ref_berman}, different formulations of Lights Out on different graphs were presented, as well as evaluations of the computational complexity of different formulations of the problem.
% \Abu{There is a paper that -- this does not sound a good way to start -- You can say In [12], there were presented ....}
%( Marlow Anderson, Todd Feil (1998). "Turning Lights Out with Linear Algebra" %https://web.archive.org/web/20140815155142/https://www.math.ksu.edu/math551/math551a.f06/lights_out.pdf ).
%Removed https://raw.org/research/solving-lightsout-using-linear-algebra/
%There is also a paper Lights Out on graphs (https://arxiv.org/pdf/1903.06942), that checks different formulations of Lights Out on different graphs, and checks the computation complexity of different formulations of the problem.

The Lights Out problem is an interesting candidate for using Grover's Search algorithm. $3 \times 3$ grid instance was considered in \cite{ref_vidwans}. As it is the original definition of the Lights Out problem, where lamp toggles are symmetric, the solution just tracks the lamps that had to be clicked in order to solve the problem for a specific initial configuration. The circuit in this article is big, and we consider smaller instances with different encoding of the problem to make our benchmarking instances.
%\Abu{bigger?}

\section{Implementation of Problem instances}

In our design, we use state 0 to mark a specific lamp as switched off, and state 1 to mark a specific lamp as switched on. For each instance, we implement an oracle that marks the lamp index to click that solves the Lights Out puzzle. Grover’s Search then operates on this space of click candidates, and the output is expected to indicate which lamp to click. In general, more lamps should be clicked to turn off all lamps, but here, as our instance problems are small, the problem admits a single-click solution.

\subsection{Lights Out instance for $2 \times 2$ grid}

%\Abu{A figure of this instance}

Given an initial configuration of 4 lamps in a $2 \times 2$ grid, the task is to find which lamp we need to click in order to make all 4 lamps switched off. On a grid, clicking a lamp changes its state, and the states of all direct neighbors (right, left, top, or bottom). In the case of a $2 \times 2$ grid, clicking specific lamp switches states of 3 lamps in total, leaving only the opposite-diagonal one unaffected.

\begin{figure}[htbp]
    \centering
%    \includegraphics[width=0.5\linewidth]{iqm_results_3.png}
    % First Subfigure
    \begin{subfigure}[b]{0.48\linewidth}
        \centering
        \includegraphics[width=0.3\linewidth]{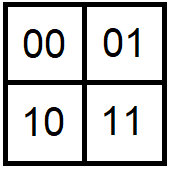}
        \caption{Grid with $2 \times 2$ lamps.}
        \label{fig0_left}
    \end{subfigure}
    \hfill % Adds space between the images
    % Second Subfigure
    \begin{subfigure}[b]{0.48\linewidth}
        \centering
        \includegraphics[width=0.8\linewidth]{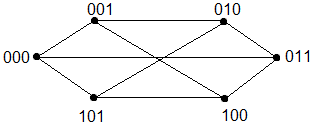}
        \caption{M\"obius ladder with 6 lamps.}
        \label{fig0_right}
    \end{subfigure}
    
    \caption{Graphs and basis states for the implementation of Lights Out instances.}
    \label{fig0}
\end{figure}

%\begin{figure}
%\centering
%\includegraphics[width=0.15\textwidth]{2x2_lamps_grid.png}
%\caption{Grid with $2 \times 2$ lamps.} \label{fig0}
%\end{figure}

In our implementation, we use 2 qubits to store the indices of the lamps to be clicked. Lamps are represented by states on the top row as 00 and 01, and on the bottom row as 10 and 11 (see Fig.~\ref{fig0_left}). We use 4 qubits to store the initial configuration of 4 lamps, as well as the updated configuration after the clicking operation. We use one qubit to compute the oracle function value, and this qubit will also flip the phase of the correct solution. We add 2 auxiliary qubits, as we implement the NOT operator controlled by 4 qubits with 3 Toffoli operations.% \Abu{the following part is confusing, not clear} and we decided to make manually this control operation with 3 Toffoli operations.% \Abu{Re-write this part by replacing the word ``need''}

\begin{figure}
\includegraphics[width=\textwidth]{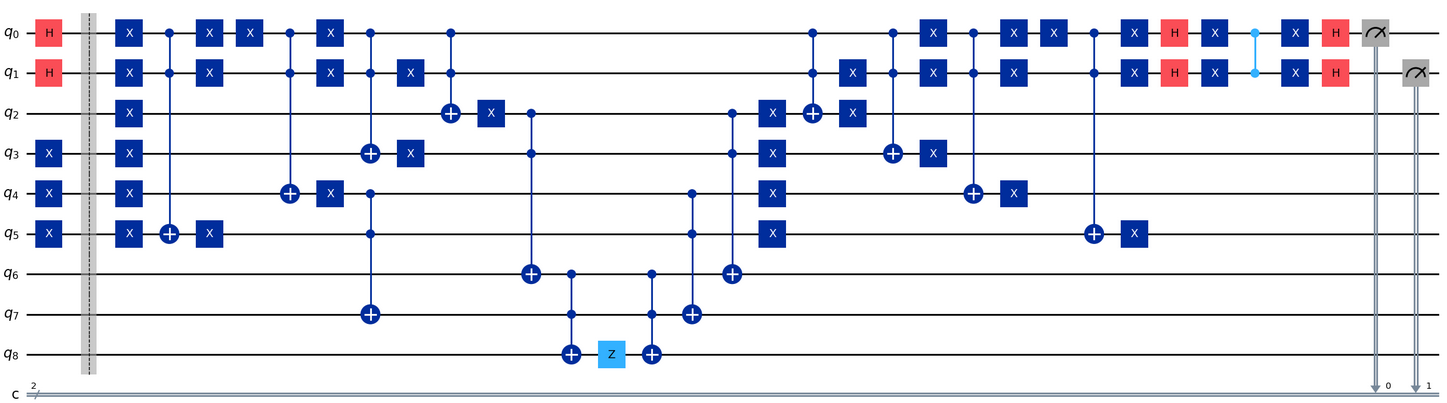}
\caption{Implementation of Grover's Search for Lights Out problem on $2 \times 2$ grid of lamps.} \label{fig1}
\end{figure}

The implementation of the circuit is depicted on
Fig.~\ref{fig1}. The qubits $q_0$ and $q_1$ store the index of the lamp to be clicked, and Grover's Search space is initialized in superposition with Hadamard operators. The qubits $q_2$, $q_3$, $q_4$, and $q_5$ store the states of the lamps. We use X-gate wherever needed before the barrier to set the initial configuration. To reduce the number of multi-qubit operations in the oracle, we remark that clicking one lamp changes the states of all lamps except one; therefore, we switch the state of all lamps with an X-gate, and switch the state again for the lamp that is not a neighbor of the clicked one. As a result, the circuit contains 14 Toffoli gates in total. The last 5 layers before the measurement operator implement Grover's Diffusion operation.

This circuit has 9 qubits and contains 14 Toffoli operations, 1 two-qubit gate (CZ), and a total depth of 31 layers (before the measurement) before any optimization or transpilation process. Since the circuit searches for 1 marked element among 4 possible candidates, we have one iteration, and Grover's Search is expected to return state \texttt{11} with probability 1.

\subsection{Lights Out instance for M\"obius ladder with 6 lamps}

%\st{Initial formulation of the structure of the lamp switching operations was as follows.}
We are given $n \geq 6$ lamps arranged in a circle, where $n$ is an even number. By clicking a lamp, the states of both neighbors and the lamp on the opposite side of the circle are changed. Therefore, if the lamps are sequentially indexed around the circle and we click a lamp indexed by $a$, then the states of the lamps indexed by $(a+1) \mod{n}$, $(a-1) \mod{n}$, and $(a+n/2) \mod{n} $ are changed. This design corresponds to non-reflexive Lights Out on a M\"obius Ladder. Each lamp is connected to three other lamps, and clicking a lamp affects only the states of direct neighbors.

%\begin{figure}
%\centering
%\includegraphics[width=0.4\textwidth]{mobius_6_lamps.png}
%\caption{M\"obius ladder with 6 lamps.} \label{fig2}
%\end{figure}

The graph structure for 6 lamps with their enumeration with basis states is shown in Fig.~\ref{fig0_right}. As we can see, clicking any lamp triggers every second lamp, whose right-most bit value is opposite to the value of the right-most bit of the clicked lamp. This observation allows us to do some simplifications. Based on the right-most bit of the clicked lamp, we can switch the states of the corresponding 3 lamps. Also, since there are no lamps with indexes 110 and 111, we added an extra check that there is no suitable element with the left-most indexes 11.

In this implementation, we use 3 qubits to store the indices of the lamps to be clicked. We use 6 qubits to store the initial configuration of our lamps, and the updated configuration after the clicking operation. We use one more qubit to store the verification that a non-existing lamp (110 or 111) has not been clicked. We use the NOT operator, controlled by 7 qubits; therefore, we add 5 more auxiliary qubits to implement control with 6 RCCX-gates (RCCX-gate serves as an efficient alternative to the Toffoli gate).

\begin{figure}
\includegraphics[width=\textwidth]{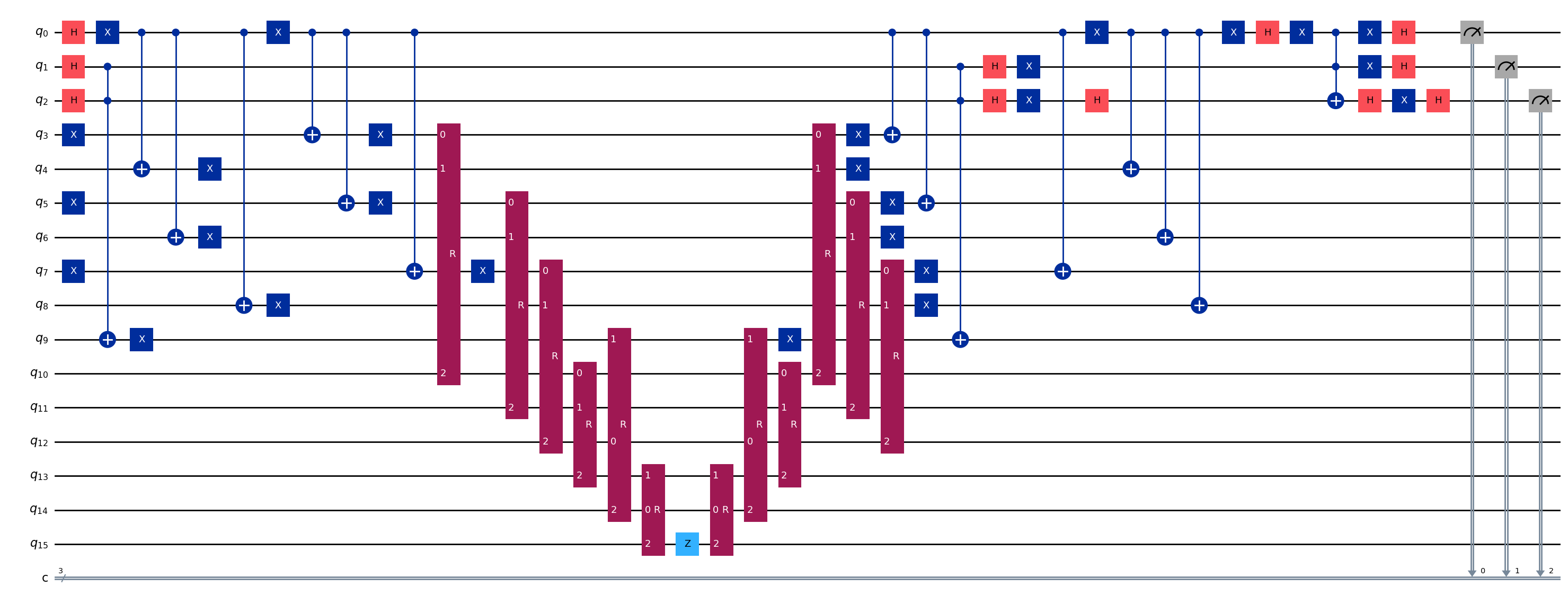}
\caption{Implementation of Grover's Search for Lights Out problem on M\"obius ladder with 6 lamps.} \label{fig3}
\end{figure}

The implementation of the circuit is depicted in Fig.~\ref{fig3}. The qubits $q_0$, $q_1$, and $q_2$ store the index of the lamp to be clicked, and as Grover's Search space is initialized in superposition with Hadamard operators. The qubits $q_3$, $q_4$, $q_5$, $q_6$, $q_7$, and $q_8$ store the states of the lamps. The initial configuration is given as input -- X-operations before the barrier on the left side. For oracle implementation, we implement two checks. First, the state of $q_0$ affects the choice of 3 lamps for which the state should be changed. Second, the states of $q_1$ and $q_2$ should not be equal to 1, and the result of this check is stored in $q_9$. The last layers of the circuit on the first 3 qubits implement Grover's Diffusion operation.
%\Abu{The second part should be re-written, it is not clear what it does mean}

This circuit has 16 qubits and contains 3 Toffoli operations, 12 RCCX-gates (in Fig.~\ref{fig3} denoted as R), 12 two-qubit gates (CNOTs), and a total depth of 32 layers (before measurement) before any optimizations and transpilation. Since the circuit searches for 3 marked elements among 8 possible candidates (we have 3 lamps out of 6 that work for the solution, but the total search space consists of 8 elements), we have one iteration, and Grover's Search is expected to return states `001', `011', or `101', each with probability approximately $0.28$.

\subsection{Supplementary baseline circuits (SAT and two-qubit sanity check)}

When we implemented the circuits for Lights Out on IQM devices, the measured distributions were near-uniform (the observed counts for the expected state stayed within $\approx 3 \sigma$ of uniform 250). To contextualize this result, we include two supplementary diagnostic baselines based on Grover's Search -- a small circuit for SAT and a smaller two-qubit Grover's Search instance. These baselines test whether the effects of Grover's search are observable on IQM devices in the case of small circuits.  %\Abu{Always use two `-'`-'}

%The circuit for the small SAT instance implements the formula $(q_0) \land (\neg q_1)$, and the expected output of the circuit is state \texttt{01} with probability 1. The circuit has 5 qubits and 3 Toffoli gates. In our experiments, one of the IQM devices provided near-uniform output, so we decided to run an even smaller circuit with Grover's Search, just to verify that everything works properly. The circuit uses just two qubits and has only two two-qubit operations. The expected output is state \texttt{10} with probability 1. Images of both circuits are publicly available at \url{https://github.com/infenrio/lights_out_quantum/sat_and_2qubits.pdf}.

%suggesting that hardware noise may dominate. so we concluded that it may be worth to try some smaller circuits. In order to test IQM devices, we constructed a circuit implementing Grover's Search for small instances of SAT problem. 

\begin{figure}
\centering
\includegraphics[width=0.6\textwidth]{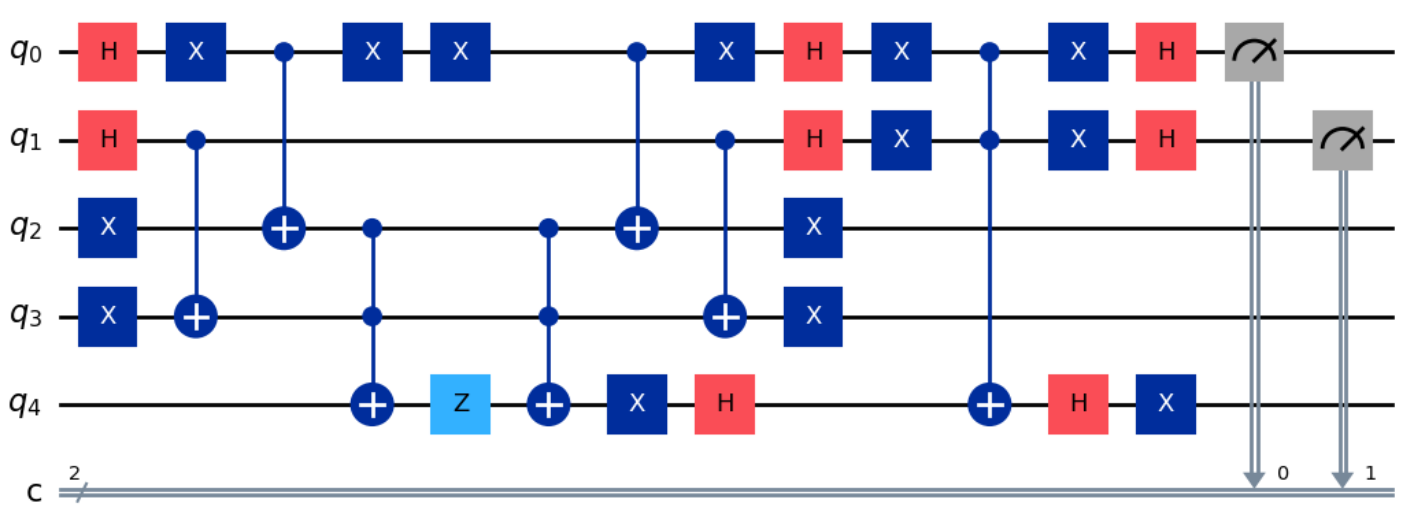}
\caption{Implementation of Grover's Search for a small SAT problem.} \label{fig4}
\end{figure}

The circuit that we prepared for the small SAT instance is shown in Fig.~\ref{fig4}. Here, $q_0$ and $q_1$ serve as the formula variables, and the SAT formula to solve is $(q_0) \land (\neg q_1)$. $q_2$ and $q_3$ store the calculated values for clauses, and $q_4$ is used to perform phase change. The expected output of the circuit is state \texttt{01} with probability 1.

\begin{figure}
\centering
\includegraphics[width=0.6\textwidth]{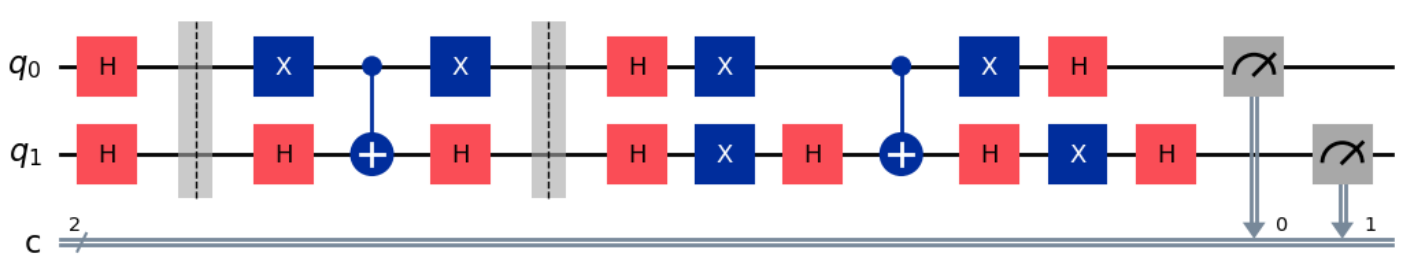}
\caption{Implementation of a small imitation of Grover's Search.} \label{fig5}
\end{figure}

In our experiments, one of the IQM devices provided near-uniform output, so we decided to run an even smaller circuit with Grover's Search, just to verify that everything works properly (see Fig.~\ref{fig5}). The circuit uses just two qubits and has only two two-qubit operations. The expected output is state \texttt{10} with probability 1.

\section{Methods for running circuits on quantum hardware}

To run the quantum circuit on a quantum computer, the circuit is first translated to the native operations of the computer. Qiskit has the option of using a transpiler that performs the translation of the circuit to map the qubits and operations of the circuit to the actual hardware architecture. The transpiler has an option to provide the \optlevel parameter, where the higher level uses more options to optimize the circuit in a way that it has lower depth, number of gates, and uses qubits and their connectivity more optimally (e.g., picking qubits that showed better results at calibration benchmarks). It is important to note that different runs of the transpiler for the same circuit with the same parameters may produce different resulting circuits (i.e., some randomness is involved in the transpilation process, and, additionally, differences in hardware calibration data also affect the choices made by the transpiler). %\Abu{Mostly there is no comma before ``that''.}

There are two options to launch the circuit in Qiskit on a quantum computer: Sampler and Estimator. When we use Sampler, the circuit with measurements is launched on a quantum computer, and we receive measurement outcomes as bitstring distributions over the total number of $\texttt{shots}$.
In the case of Estimator, a quantum computer receives a circuit and observables, and produces expectation values and uncertainty (standard deviations). While Estimator is designed more for tasks related to the expected value of the Hamiltonians or cost operators, it supports error mitigation techniques that can be controlled through the \reslevel parameter. %\Abu{perhaps all parameters in italic font?}

It is important to note that the Estimator with a high \reslevel consumes significantly more runtime on IBM hardware than the Sampler. This affected the number of experiments that we performed during this research.

\section{Setup for experiments}

The first step is the preparation of the circuits that we described in Section 3. Each prepared circuit is tested on a classical simulator (AerSimulator) that imitates a noiseless quantum computer. This allows us to ensure that the circuit is properly prepared, as well as verify the ideal (simulated) output probabilities, and compare them with measured shot-based frequencies on hardware.
%the expected outcomes with related probabilities, which are mentioned in the previous section.

%\Abu{You can define a variable for all these parameters and real quantum hardware name, etc. to easily refer them and make changes globally later.}

Then, we started our benchmarking process. We started with the circuit for a $2 \times 2$ grid of lamps instance and prepared it for quantum computers, available at the IBM platform: \marrakesh, \fez, and \torino. The circuit was launched on each device in 5 different modes. We used Qiskit code with connection API to send circuits, transpiled with \optlevel values 0, 1, 2, and 3. We used the Sampler method. The 5th mode that we tried was to transform the circuit into a diagram and launch directly from the IBM composer from the IBM platform. We recorded the calibration data of computers and the data about the circuits that were actually launched on the devices after the transpilation process. We saved the outcomes of the circuit executions. To verify the consistency of the outcomes, we also did a sequence of 5-day tests, where we launched the circuit on all 3 IBM devices with transpilation with \optlevel 2, and with IBM platform Composer.

For IQM, we launched the circuit for a $2 \times 2$ grid of lamps instance on all available devices: IQM Emerald, IQM Garnet, and IQM Sirius. We used Qiskit code with connection API to send circuits, transpiled with \optlevel values 0, 1, 2, and 3. The results were near-uniform outputs; therefore, we added supplementary baseline Grover's Search circuits (SAT and a two-qubit sanity check) to interpret the Lights Out results for IQM devices.

% Therefore, for the next steps, we continued the Lights Out benchmarking on IBM devices, and we added supplementary baseline Grover's Search circuits (SAT and a 2-qubit sanity check) to interpret the Lights Out results for IQM devices.

%we decided to launch larger circuit on IBM devices, and smaller circuits on IQM devices.

For IBM devices, we started the implementation of the circuit for 6 lamps on M\"obius ladder. We did several iterations trying to optimize and simplify the circuit until we reached the solution that is on Fig.~\ref{fig3}. For each iteration, we tried to run the circuit with \optlevel 3 transpilation on the device that showed best performance with the $2 \times 2$ grid circuit. In each iteration, the results were near-uniform, so we kept improving the circuit. We launched the final circuit on all 3 IBM devices with Sampler and \optlevel 3, with Estimator for \reslevel values 0, 1, and 2, and in IBM platform's circuit composer. We note that the experiments on Composer were launched on a different day than the other experiments. We recorded the calibration data, actual circuits that were launched, and outcomes. %We note that in hopes of better performance for implementations for Sampler and Estimator, we replaced all Toffoli gates, except the first two and the last one, with RCCX gates, while in Composer we used the circuit as it is shown in Fig.~\ref{fig2}.
We also launched the circuit for 6 lamps on M\"obius ladder on IQM devices with Sampler and \optlevel 3 and 2, and recorded the transpilation results and the outcomes of the experiments.

Then, for IQM devices, we launched the circuit for a small SAT problem on all 3 devices, and recorded the calibration data and outcomes of the experiments. After seeing that in some cases the outputs were near-uniform, we decided to run the smallest Grover's Search implementation on 2 qubits on all devices, and saved all the related data of the experiments. For additional context, we also launched the circuit for a small SAT problem on IBM devices.

For IBM devices, we launched the circuits with parameter $\texttt{shots}$ set to 4000; for IQM devices, parameter $\texttt{shots}$ was set to 1000 (because the available runtime quota is smaller, so we saved computational time). The sampling uncertainty ($1 \sigma$) for a reported frequency $\hat{p}$ from $N$ shots is $\sqrt{\hat{p}(1-\hat{p})/N}$, with the worst-case bound $\leq1/(2\sqrt{N})$, i.e., $\leq 0.0079$ for $N=4000$ and $\leq 0.0158$ for $N=1000$.

\section{Results}

\subsection{Lights Out on $2 \times 2$ grid}

We summarize the results of our experiments on IBM devices in Table~\ref{tab:circuit_results_grid}, and the statistics of the circuits that were actually run on hardware in Table~\ref{tab:circuit_stats_grid}.

\begin{table}[t]
\caption{Measured output distributions (4000 $\texttt{shots}$ each) for the circuit for a $2 \times 2$ grid of lamps. The first line contains the correct number of counts for the correct state \texttt{11}, together with the relative frequency, second line contains the remaining counts. The data is shown for 5 different launch modes -- IBM platform's composer, and Sampler with \optlevel value set to 3, 2, 1, and 0.}
\label{tab:circuit_results_grid}
\centering
\small
\setlength{\tabcolsep}{4pt}
\renewcommand{\arraystretch}{1.15}
\newcommand{\wrong}[1]{{\fontsize{6}{7}\selectfont #1}}

\begin{tabular}{lccc}
\toprule
Mode & \textbf{ibm\_marrakesh} & \textbf{ibm\_fez} & \textbf{ibm\_torino} \\
\midrule
Composer &
\makecell[l]{\textbf{11: 2304 (0.576)}\\ \wrong{00: 515 \; 01: 665 \; 10: 516}} &
\makecell[l]{\textbf{11: 1645 (0.411)}\\ \wrong{00: 759 \; 01: 828 \; 10: 768}} &
\makecell[l]{\textbf{11: 954 (0.239)}\\ \wrong{00: 1014 \; 01: 950 \; 10: 1082}} \\
%\addlinespace

opt=3 &
\makecell[l]{\textbf{11: 2018 (0.505)}\\ \wrong{00: 563 \; 01: 716 \; 10: 703}} &
\makecell[l]{\textbf{11: 1417 (0.354)}\\ \wrong{00: 853 \; 01: 933 \; 10: 797}} &
\makecell[l]{\textbf{11: 1796 (0.449)}\\ \wrong{00: 599 \; 01: 896 \; 10: 709}} \\
opt=2 &
\makecell[l]{\textbf{11: 1896 (0.474)}\\ \wrong{00: 621 \; 01: 731 \; 10: 752}} &
\makecell[l]{\textbf{11: 1528 (0.382)}\\ \wrong{00: 842 \; 01: 814 \; 10: 816}} &
\makecell[l]{\textbf{11: 1633 (0.408)}\\ \wrong{00: 709 \; 01: 858 \; 10: 800}} \\
opt=1 &
\makecell[l]{\textbf{11: 1780 (0.445)}\\ \wrong{00: 644 \; 01: 868 \; 10: 708}} &
\makecell[l]{\textbf{11: 1460 (0.365)}\\ \wrong{00: 718 \; 01: 865 \; 10: 957}} &
\makecell[l]{\textbf{11: 1553 (0.388)}\\ \wrong{00: 754 \; 01: 934 \; 10: 759}} \\
opt=0 &
\makecell[l]{\textbf{11: 1370 (0.343)}\\ \wrong{00: 749 \; 01: 1090 \; 10: 791}} &
\makecell[l]{\textbf{11: 1035 (0.259)}\\ \wrong{00: 937 \; 01: 1097 \; 10: 931}} &
\makecell[l]{\textbf{11: 1081 (0.270)}\\ \wrong{00: 942 \; 01: 1091 \; 10: 886}} \\
\bottomrule
\end{tabular}
\end{table}

\begin{table}[t]
\caption{Statistics for post-transpilation circuit for $2 \times 2$ grid of lamps  (all circuits use 9 active qubits). The data is shown for 5 different launch modes -- IBM platform's composer, and Sampler with \optlevel value set to 3, 2, 1, and 0.}
\label{tab:circuit_stats_grid}
\centering
\small
\setlength{\tabcolsep}{3.5pt}      % tighter columns (optional)
\renewcommand{\arraystretch}{1.05} % slightly taller rows (optional)

\begin{tabular}{r rrr rrr rrr}
\toprule
& \multicolumn{3}{c}{marrakesh} & \multicolumn{3}{c}{fez} & \multicolumn{3}{c}{torino} \\
\cmidrule(lr){2-4}\cmidrule(lr){5-7}\cmidrule(lr){8-10}
Mode & 2q op & depth & size & 2q op & depth & size & 2q op & depth & size \\
\midrule
composer & 170 & 293 & 633 & 167 & 295 & 631 & 173 & 341 & 650 \\
opt=3 & 165 & 397 & 714 & 173 & 437 & 754 & 170 & 430 & 720 \\
opt=2 & 175 & 405 & 756 & 177 & 408 & 749 & 175 & 399 & 747 \\
opt=1 & 217 & 467 & 932 & 229 & 554 & 962 & 199 & 455 & 882 \\
opt=0 & 328 & 715 & 1560 & 376 & 774 & 1704 & 310 & 712 & 1506 \\
%\bottomrule
\bottomrule
\end{tabular}
\end{table}

The results of running the circuits on IQM devices are summarized in Table~\ref{tab:iqm_results_grid}, and the statistics of the circuits that were actually run on hardware in Table~\ref{tab:iqm_circuit_stats_grid}.
%\begin{itemize}
%    \item Emerald: \{'00': 255, '01': 238, '10': 246, '11': 261\}
%    \item Garnet: \{'00': 233, '01': 232, '10': 261, '11': 274\}
%    \item Sirius: \{'00': 199, '01': 253, '10': 294, '11': 254\}
%\end{itemize}

%\begin{figure}
%\centering
%\includegraphics[width=0.6\textwidth]{iqm_results_1.png}
%\caption{Measured output distributions (1000 $\texttt{shots}$ each) for the circuit for $2 \times 2$ grid of lamps on IQM devices. Expected output is state \texttt{11}.} \label{fig_iqm_1}
%\end{figure}

\begin{table}[t]
\caption{Measured counts for the correct state \texttt{11} (out of 1000 $\texttt{shots}$) for the circuit for a $2 \times 2$ grid of lamps. The data is shown for 4 different launch modes -- Sampler with \optlevel value set to 3, 2, 1, and 0.}
\label{tab:iqm_results_grid}
\centering
\small
\setlength{\tabcolsep}{4pt}
\renewcommand{\arraystretch}{1.15}
\newcommand{\wrong}[1]{{\fontsize{6}{7}\selectfont #1}}

\begin{tabular}{lccc}
\toprule
Mode & \textbf{Emerald} & \textbf{Garnet} & \textbf{Sirius} \\
\midrule

opt=3 & 266 & 258 & 277 \\
opt=2 & 218 & 230 & 269 \\
opt=1 & 246 & 289 & 277 \\
opt=0 & 265 & 284 & 242 \\
\bottomrule
\end{tabular}
\end{table}

\begin{table}[t]
\caption{Statistics for post-transpilation circuit for $2 \times 2$ grid of lamps. The data is shown for 4 different launch modes -- Sampler with \optlevel value set to 3, 2, 1, and 0.}
\label{tab:iqm_circuit_stats_grid}
\centering
\small
\setlength{\tabcolsep}{3.2pt}      % tighter columns (optional)
\renewcommand{\arraystretch}{1.05} % slightly taller rows (optional)

\begin{tabular}{r rrrr rrrr rrrr}
\toprule
& \multicolumn{4}{c}{Emerald} & \multicolumn{4}{c}{Garnet} & \multicolumn{4}{c}{Sirius} \\
\cmidrule(lr){2-5}\cmidrule(lr){6-9}\cmidrule(lr){10-13}
Mode & qub & 2q op & depth & size & qub & 2q op & depth & size & qub & 2q op & depth & size \\
\midrule
opt=3 & 9 & 151 & 246 & 403 & 9 & 151 & 232 & 402 & 10 & 202 & 221 & 322 \\
opt=2 & 9 & 152 & 230 & 392 & 9 & 145 & 227 & 397 & 10 & 202 & 221 & 322 \\
opt=1 & 11 & 166 & 246 & 418 & 10 & 178 & 257 & 454 & 10 & 197 & 215 & 315 \\
opt=0 & 12 & 244 & 673 & 1303 & 9 & 211 & 694 & 1204 & 10 & 227 & 582 & 968 \\
%\bottomrule
\bottomrule
\end{tabular}
\end{table}

We did more runs on IBM devices to check the consistency of the results. We ran the circuit for a $2 \times 2$ grid on all 3 IBM devices, with the Qiskit transpiler set to \optlevel 2, and with IBM platform composer (see Table~\ref{tab:circuit_results_5_days}).

%\begin{table}[t]
%\caption{Number of counts, when the correct state \texttt{11} was measured. All runs use 4000 shots.}
%\label{tab:circuit_results_5_days}
%\centering
%\small
%\setlength{\tabcolsep}{3.5pt}
%\renewcommand{\arraystretch}{1.2}
%
%\begin{tabular}{lcccccccccc}
%\toprule
%Device &
%\makecell[c]{D\_1\\opt} &
%\makecell[c]{D\_1\\comp} &
%\makecell[c]{D\_2\\opt} &
%\makecell[c]{D\_2\\comp} &
%\makecell[c]{D\_3\\opt} &
%\makecell[c]{D\_3\\comp} &
%\makecell[c]{D\_4\\opt} &
%\makecell[c]{D\_4\\comp} &
%\makecell[c]{D\_5\\opt} &
%\makecell[c]{D\_5\\comp}  \\
%\midrule

%marrakesh &
%$1757$ &
%$2132$ &
%$908$ &
%$862$ &
%$1774$ &
%$1111$ &
%$1794$ &
%$1246$ &
%$1811$ &
%$1427$ \\

%fez &
%$1566$ &
%$1234$ &
%$1719$ &
%$2022$ &
%$1823$ &
%$1568$ &
%$1526$ &
%$1136$ &
%$1112$ &
%$1329$ \\

%torino &
%$1799$ &
%$1026$ &
%$1328$ &
%$1622$ &
%$1746$ &
%$1603$ &
%$1325$ &
%$959$ &
%$1182$ &
%$1037$ \\

%\bottomrule
%\end{tabular}
%\end{table}

\begin{table}[t]
\caption{Observed frequencies of the correct state \texttt{11} for the circuit for $2 \times 2$ grid of lamps. The experiments were run for 5 consecutive days ($D\_x$ denotes Day $x$), and for two different modes -- Sampler with \optlevel 2 and IBM platform's composer.}
\label{tab:circuit_results_5_days}
\centering
\small
\setlength{\tabcolsep}{3pt}
\renewcommand{\arraystretch}{1.2}

\begin{tabular}{lcccccccccc}
\toprule
Device &
\makecell[c]{D\_1\\opt} &
\makecell[c]{D\_1\\comp} &
\makecell[c]{D\_2\\opt} &
\makecell[c]{D\_2\\comp} &
\makecell[c]{D\_3\\opt} &
\makecell[c]{D\_3\\comp} &
\makecell[c]{D\_4\\opt} &
\makecell[c]{D\_4\\comp} &
\makecell[c]{D\_5\\opt} &
\makecell[c]{D\_5\\comp}  \\
\midrule

marrakesh &
$0.440$ &
$0.533$ &
$0.227$ &
$0.216$ &
$0.444$ &
$0.278$ &
$0.449$ &
$0.312$ &
$0.453$ &
$0.357$ \\

fez &
$0.392$ &
$0.309$ &
$0.430$ &
$0.506$ &
$0.456$ &
$0.392$ &
$0.382$ &
$0.284$ &
$0.278$ &
$0.332$ \\

torino &
$0.450$ &
$0.257$ &
$0.332$ &
$0.406$ &
$0.437$ &
$0.401$ &
$0.331$ &
$0.240$ &
$0.296$ &
$0.259$ \\

\bottomrule
\end{tabular}
\end{table}

\subsection{Lights Out on M\"obius ladder with 6 lamps}

We summarize the results of our experiments on IBM devices in Table~\ref{tab:circuit_results_mobius}, and statistics of the circuits that were actually run on hardware in Table~\ref{tab:circuit_stats_mobius}. All final circuits on hardware had 16 qubits, except that the composer generated on \marrakesh~circuit with 19 qubits, and on \torino~circuit with 17 qubits.

\newcommand{\subcell}[1]{{\scriptsize #1}} % make the 2nd line smaller

\begin{table}[t]
\caption{Total success estimate for the three correct outputs \texttt{001}, \texttt{011}, \texttt{101}
(ideal: $\approx 0.28$ each; total: $\approx 0.84$) for the circuit for 6 lamps. Experiments were performed in 5 modes -- Sampler with \optlevel 3, Estimator with \reslevel 0, 1, and 2, and IBM platform's composer. For Estimator, we report the probability estimate returned by the primitive (mean $\pm$ reported uncertainty).}
\label{tab:circuit_results_mobius}
\centering
\small
\setlength{\tabcolsep}{3.5pt}
\renewcommand{\arraystretch}{1.2}

\begin{tabular}{lccccc}
\toprule
Device &
\makecell[c]{Sampler\\(opt=3)} &
\makecell[c]{Estimator\\($r{=}0$)} &
\makecell[c]{Estimator\\($r{=}1$)} &
\makecell[c]{Estimator\\($r{=}2$)} &
Comp. \\
\midrule

marrakesh &
%\makecell[c]{\textbf{0.4355}\\ \subcell{001/011/101: 612/591/539}} &
$0.4355$ &
$0.4637 \pm 0.0076$ &
$0.4809 \pm 0.0085$ &
$0.5358 \pm 0.0957$ &
$0.3753$ \\
%0.4335$ \\
%\makecell[c]{\textbf{0.4335}\\ \subcell{001/011/101: 548/605/581}} \\

fez &
%\makecell[c]{\textbf{0.390}\\ \subcell{001/011/101: 491/554/515}} &
$0.390$ &
$0.3398 \pm 0.0076$ &
$0.3423 \pm 0.0074$ &
$0.4280 \pm 0.1223$ &
$0.3345$ \\
%$0.3378$ \\
%\makecell[c]{\textbf{0.3378}\\ \subcell{001/011/101: 445/477/429}} \\

torino &
%\makecell[c]{\textbf{0.374}\\ \subcell{001/011/101: 490/510/496}} &
$0.374$ &
$0.4293 \pm 0.0075$ &
$0.4307 \pm 0.0074$ &
$0.4594 \pm 0.0279$ &
$0.373$ \\
%$0.4478$ \\
%\makecell[c]{\textbf{0.4478}\\ \subcell{001/011/101: 577/679/535}} \\

\bottomrule
\end{tabular}
\end{table}

\begin{table}[t]
\caption{The statistics for the post-transpilation circuit for 6 lamps. The data is shown for 3 different launch modes -- Sampler with \optlevel 3, Estimator with \reslevel 2, and IBM platform's composer.}
\label{tab:circuit_stats_mobius}
\centering
\small
\setlength{\tabcolsep}{3.5pt}      % tighter columns (optional)
\renewcommand{\arraystretch}{1.05} % slightly taller rows (optional)

\begin{tabular}{r rrr rrr rrr}
\toprule
& \multicolumn{3}{c}{marrakesh} & \multicolumn{3}{c}{fez} & \multicolumn{3}{c}{torino} \\
\cmidrule(lr){2-4}\cmidrule(lr){5-7}\cmidrule(lr){8-10}
method & 2q op & depth & size & 2q op & depth & size & 2q op & depth & size \\
\midrule
opt=3 & 191 & 215 & 802 & 191 & 213 & 796 & 203 & 234 & 798 \\
res=2 & 191 & 214 & 799 & 191 & 212 & 793 & 203 & 228 & 790 \\
%composer & 264 & 313 & 951 & 249 & 325 & 913 & 258 & 320 & 936 \\
composer & 243 & 227 & 748 & 219 & 206 & 685 & 234 & 257 & 739 \\
\bottomrule
\end{tabular}
\end{table}

The results of running the circuits on IQM devices are summarized in Table~\ref{tab:iqm_results_mobius}, and the statistics of the circuits that were actually run on hardware in Table~\ref{tab:iqm_circuit_stats_mobius}.

\begin{table}[t]
\caption{Measured output distributions (1000 $\texttt{shots}$ each) for the circuit for 6 lamps. The first line contains the total number of counts for the expected states \texttt{001}, \texttt{011}, \texttt{101}, together with the relative frequency; the second line contains the counts for the expected states. The data is shown for 2 different launch modes -- Sampler with \optlevel value set to 3 and 2.}
\label{tab:iqm_results_mobius}
\centering
\small
\setlength{\tabcolsep}{4pt}
\renewcommand{\arraystretch}{1.15}
\newcommand{\wrong}[1]{{\fontsize{6}{7}\selectfont #1}}

\begin{tabular}{lccc}
\toprule
Mode & \textbf{Emerald} & \textbf{Garnet} & \textbf{Sirius} \\
\midrule

opt=3 &
\makecell[l]{\textbf{total: 441 (0.441)}\\ \wrong{001: 191 \; 011: 148 \; 101: 102}} &
\makecell[l]{\textbf{total: 397 (0.397)}\\ \wrong{001: 111 \; 011: 198 \; 101: 88}} &
\makecell[l]{\textbf{total: 465 (0.465)}\\ \wrong{001: 235 \; 011: 148 \; 101: 82}} \\
opt=2 &
\makecell[l]{\textbf{total: 346 (0.346)}\\ \wrong{001: 113 \; 011: 130 \; 101: 103}} &
\makecell[l]{\textbf{total: 305 (0.305)}\\ \wrong{001: 110 \; 011: 96 \; 101: 99}} &
\makecell[l]{\textbf{total: 475 (0.475)}\\ \wrong{001: 238 \; 011: 153 \; 101: 84}} \\
\bottomrule
\end{tabular}
\end{table}

\begin{table}[t]
\caption{Statistics for post-transpilation circuit for 6 lamps. The data is shown for 2 different launch modes -- Sampler with \optlevel value set to 3 and 2.}
\label{tab:iqm_circuit_stats_mobius}
\centering
\small
\setlength{\tabcolsep}{3.2pt}      % tighter columns (optional)
\renewcommand{\arraystretch}{1.05} % slightly taller rows (optional)

\begin{tabular}{r rrrr rrrr rrrr}
\toprule
& \multicolumn{4}{c}{Emerald} & \multicolumn{4}{c}{Garnet} & \multicolumn{4}{c}{Sirius} \\
\cmidrule(lr){2-5}\cmidrule(lr){6-9}\cmidrule(lr){10-13}
Mode & qub & 2q op & depth & size & qub & 2q op & depth & size & qub & 2q op & depth & size \\
\midrule
opt=3 & 17 & 106 & 112 & 294 & 16 & 104 & 102 & 285 & 17 & 169 & 179 & 271 \\
opt=2 & 16 & 109 & 103 & 304 & 17 & 123 & 106 & 333 & 17 & 169 & 179 & 271 \\
%\bottomrule
\bottomrule
\end{tabular}
\end{table}

\subsection{Supplementary baseline circuits (SAT and two-qubit sanity check)}

% Preamble:
% \usepackage{booktabs}
% \usepackage{makecell}

\newcommand{\wrong}[1]{{\scriptsize #1}} % smaller font for wrong outcomes

%To contextualize the near-uniform Lights Out results on IQM, we summarize the results of running the SAT circuit. For Sampler with \optlevel 3, the relative frequency for the correct state \texttt{01} was $0.409$ on Emerald, $0.630$ on Garnet, $0.258$ on Sirius, $0.860$ on \marrakesh, $0.790$ on \fez, and $0.835$ on \torino. For Sampler with \optlevel 0, the relative frequency for the correct state \texttt{01} was $0.241$ on Emerald, $0.509$ on Garnet, $0.260$ on Sirius, $0.484$ on \marrakesh, $0.242$ on \fez, and $0.548$ on \torino. On the two-qubit Grover's Search circuit, the relative frequency for the correct state \texttt{10} was $0.907$ on Emerald, $0.934$ on Garnet, and $0.707$ on Sirius. Detailed results are publicly available at \url{https://github.com/infenrio/lights_out_quantum/sat_and_2qubits.pdf}.

\begin{figure}[htbp]
    \centering
    \includegraphics[width=0.5\linewidth]{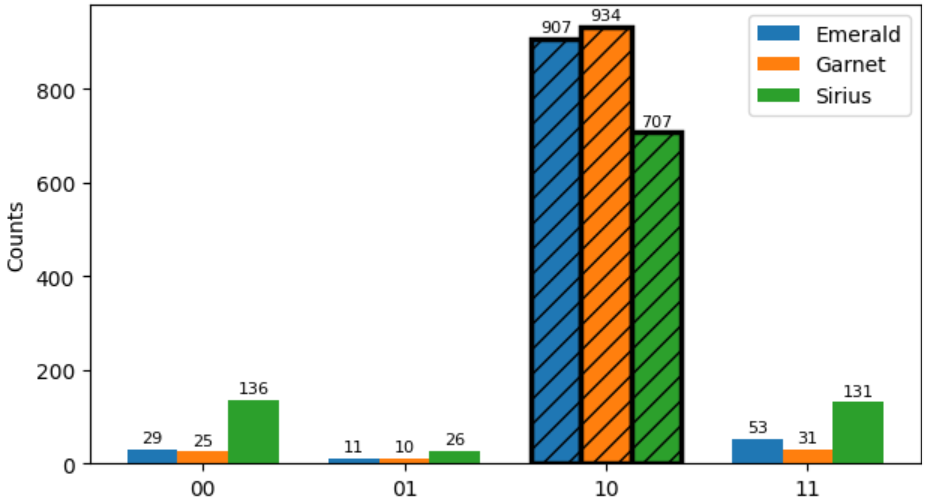}
    % First Subfigure
%    \begin{subfigure}[b]{0.48\linewidth}
%        \includegraphics[width=\linewidth]{iqm_results_2.png}
%        \caption{Results for 5-qubit SAT circuit.}
%        \label{fig:left}
%    \end{subfigure}
%    \hfill % Adds space between the images
    % Second Subfigure
%    \begin{subfigure}[b]{0.48\linewidth}
%        \includegraphics[width=\linewidth]{iqm_results_3.png}
%        \caption{Results for 2-qubit circuit.}
%        \label{fig:right}
%    \end{subfigure}
    
    \caption{Measured output distributions (1000 $\texttt{shots}$ each) on IQM devices for supplementary diagnostic baseline 2-qubit circuit. Expected output is state \texttt{10}.}
    \label{fig_iqm_2}
\end{figure}

To contextualize the near-uniform Lights Out results on IQM, we summarize the results of running the SAT circuit in Table~\ref{tab:01_success_iqm} and the small Grover's Search circuit in Fig.~\ref{fig_iqm_2} on IQM devices, and on IBM devices in Table~\ref{tab:01_success_ibm}.

%The results of running the SAT circuit on IQM devices are the following:
%\begin{itemize}
%    \item Emerald: \{'00': 177, '01': \textbf{375}, '10': 197, '11': 251\}
%    \item Garnet: \{'00': 136, '01': \textbf{583}, '10': 141, '11': 140\}
%    \item Sirius: \{'00': 200, '01': \textbf{264}, '10': 245, '11': 291\}
%\end{itemize}

%The results of running the small Grover's Search circuit on IQM devices are the following:
%\begin{itemize}
%    \item Emerald: \{'00': 29, '01': 11, '10': \textbf{907}, '11': 53\}
%    \item Garnet: \{'00': 25, '01': 10, '10': \textbf{934}, '11': 31\}
%    \item Sirius: \{'00': 136, '01': 26, '10': \textbf{707}, '11': 131\}
%\end{itemize}

\begin{table}[t]
\caption{Measured output distributions (1000 $\texttt{shots}$ each) for the SAT baseline circuit. The first line contains the number of counts for the correct state \texttt{01}, together with relative frequency, and the second line contains the remaining counts. The data is shown for Sampler mode with \optlevel 3 and 0.}
\label{tab:01_success_iqm}
\centering
\small
\setlength{\tabcolsep}{3.5pt}
\renewcommand{\arraystretch}{1.15}

\begin{tabular}{lcc}
\toprule
Device & \makecell[c]{opt=3} & \makecell[c]{opt=0} \\
\midrule
Emerald &
\makecell[c]{\textbf{01: 409 (0.409)}\\\wrong{00: 148 \; 10: 135 \; 11: 308}} &
\makecell[c]{\textbf{01: 241 (0.241)}\\\wrong{00: 210 \; 10: 264 \; 11: 285}}  \\
Garnet &
\makecell[c]{\textbf{01: 630 (0.630)}\\\wrong{00: 105 \; 10: 68 \; 11: 197}} &
\makecell[c]{\textbf{01: 509 (0.509)}\\\wrong{00: 128 \; 10: 190 \; 11: 173}}  \\
Sirius &
\makecell[c]{\textbf{01: 258 (0.258)}\\\wrong{00: 250 \; 10: 237 \; 11: 255}} &
\makecell[c]{\textbf{01: 260 (0.260)}\\\wrong{00: 263 \; 10: 255 \; 11: 222}}  \\
\bottomrule
\end{tabular}
\end{table}

\begin{table}[t]
\caption{Measured output distributions (4000 $\texttt{shots}$ each) for the SAT baseline circuit. The first line contains the number of counts for the correct state \texttt{01}, together with relative frequency, and the second line contains the remaining counts. The data is shown for Sampler mode with \optlevel 3 and 0.}
\label{tab:01_success_ibm}
\centering
\small
\setlength{\tabcolsep}{3.5pt}
\renewcommand{\arraystretch}{1.15}

\begin{tabular}{lcc}
\toprule
Device & \makecell[c]{opt=3} & \makecell[c]{opt=0} \\
\midrule
ibm\_marrakesh &
\makecell[c]{\textbf{01: 3438 (0.860)}\\\wrong{00: 201 \; 10: 123 \; 11: 238}} &
\makecell[c]{\textbf{01: 1935 (0.484)}\\\wrong{00: 1084 \; 10: 352 \; 11: 629}}  \\
ibm\_fez &
\makecell[c]{\textbf{01: 3161 (0.790)}\\\wrong{00: 310 \; 10: 191 \; 11: 338}} &
\makecell[c]{\textbf{01: 966 (0.242)}\\\wrong{00: 1791 \; 10: 686 \; 11: 557}}  \\
ibm\_torino &
\makecell[c]{\textbf{01: 3339 (0.835)}\\\wrong{00: 259 \; 10: 157 \; 11: 245}} &
\makecell[c]{\textbf{01: 2193 (0.548)}\\\wrong{00: 735 \; 10: 414 \; 11: 658}}  \\
\bottomrule
\end{tabular}
\end{table}

\section{Result analysis and discussion}

In November 2025, we noticed that IBM's free plan offers new quantum computers. Initially, we tested our $2 \times 2$ grid circuit with Composer, and tested cases, where expected outcomes were $00$ and $11$. We noticed a clear pattern that \marrakesh~provided significantly better results than the other two devices, and \fez~showed better results than \torino~(its output was near-uniform). This makes sense after seeing the calibration benchmarks, and also because \torino~has Heron r1 architecture (year 2023), and the other two devices have Heron r2 architecture (year 2024). When we launched our experiments with Qiskit code, the results started to differ. Table~\ref{tab:circuit_results_grid} demonstrates that the advantage of devices may differ because of different transpilation settings. For example, with transpilation and \optlevel 3, 2, 1, \fez~performs worse than the older device \torino.

To analyze the performance of IBM devices on the circuit for a $2 \times 2$ grid, we decided to run experiments for five consecutive days to observe the difference between calibrations. As we can see from Table~\ref{tab:circuit_results_5_days}, not only do different calibrations change the clear winner, but also different circuit launch setups provide benefits for devices inconsistently. For example, we can see that on day 3, \marrakesh~performed well with Qiskit transpiler, but performed poorly with composer. Both circuits were run on different qubits, and this could have affected the difference in performance. As the average from these 10 experiments, \marrakesh~and \fez~had observed success frequency of 0.37, while \torino~had 0.34. %\Abu{probability versus frequencies??}

We paid close attention to the transpiled circuits that actually were run on IBM devices. For different runs, the circuit statistics were close to the ones in Table~\ref{tab:circuit_stats_grid}. IBM states that the default value for \optlevel of the transpiler is 2. We can see that the circuit statistics for the composer and transpiler with \optlevel values 3 and 2 are quite similar. It is worth noting that the composer uses RZZ gates that are currently used in IBM at an experimental stage. This can explain why, within the same calibration, the effect of transpilation through composer launch on different IBM devices is different. Since we see that the structure and statistics are similar among different IBM devices, we can conclude that the generated circuit was not the main contributor to the difference in performance of IBM devices. It is interesting to observe that current hardware is sometimes capable of executing a circuit on 9 qubits with more than 300 two-qubit operations, and depth exceeding 700.% \Abu{re-write here: , depth exceeding 700}. \Abu{Probably you should not use ever ``nice'' in the academic context :-), perhaps ``interesting''}

IQM devices did not demonstrate the capacity needed to successfully solve our instance of Lights Out on a $2 \times 2$ grid of lamps, as the observed frequency of the correct outcome was near-uniform. To interpret this outcome and check how devices work with Grover's Search on smaller circuits, we ran the supplementary baseline circuits. First, we tested the SAT instance with a circuit on 5 qubits. The observed success frequency on IQM Garnet was bigger than on IQM Emerald, and on IQM Emerald was bigger than on IQM Sirius, whose outcome was near-uniform. The smaller Grover's Search instance with 2 qubits confirmed the trend, with IQM Sirius providing a non-uniform outcome this time. Garnet and Emerald have similar connectivity architecture (grid), while Garnet has a smaller number of qubits.
%The observed success frequency on IQM Garnet was 0.583, IQM Emerald had a success frequency of 0.375, and the result of IQM Sirius was close to a random outcome (0.264). The smaller Grover's Search instance with 2 qubits confirmed the trend -- IQM Garnet had a success frequency of 0.934, while IQM Emerald had 0.907, and IQM Sirius had 0.707. Garnet and Emerald have similar connectivity architecture (grid), while Garnet has a smaller number of qubits. %\Abu{rewrite to remove ``-''s}

The circuit for Lights Out on M\"obius ladder was expected to be a harder challenge for IBM devices. Since in our initial experiments \marrakesh~showed the best performance, we aimed to reach successful results on that device first. We concluded that the Estimator with available resilience solutions (built-in error mitigation and error suppression) may provide satisfactory results. Table~\ref{tab:circuit_results_mobius} demonstrated the case, when \torino~performs better than \fez, while the only \marrakesh~with the strongest Estimator settings was capable of barely exceeding the observed success frequency of $0.5$. Although the quick look at Table~\ref{tab:circuit_stats_mobius} showed that statistically the circuit should not be harder than one for a $2 \times 2$ grid, the circuit involves 16 qubits compared to 9 qubits in the case of $2 \times 2$ grid, as well as success requires a correct bitstring across more measured qubits, so readout errors and decoherence accumulate more strongly. Limited hardware connectivity can force additional routing overhead, which increases the effective two-qubit error even when the nominal depth appears comparable. These effects scale unfavorably with the number of active qubits and can dominate over depth alone.
%(it has a smaller circuit depth than a $2 \times 2$ circuit, and a comparable number of two-qubit operations), with both Estimator and Sampler generating circuits of similar complexity. We should keep in mind that the main difference is that this experiment involves 16 qubits, compared to 9 qubits in $2 \times 2$ grid experiments.
%In recent years, quantum computers have reached hundreds of qubits, and there have been more concerns about the too high error rate of quantum gates, which does not allow for the full potential of hundreds of qubits. It is interesting to observe here that the number of qubits involved was the reason for higher error rates, not the total number of gates or circuit depth.
%Connectivity can be the key here, and similar observations were noticed in \cite{ref_joshi}, where authors took a detailed look at how a quantum computer’s qubit connectivity graph and noise influence Grover’s Search algorithm. %\Abu{rewrite: authors took a detailed look at}
Similar observations were noticed in \cite{ref_joshi}, where authors took a detailed look at how a quantum computer’s qubit connectivity graph and noise influence Grover’s Search algorithm.

For IQM devices, it was interesting to observe the effect of the qubit connectivity on the transpilation for the circuit for Lights Out on M\"obius ladder. The transpiled circuit was consistently smaller than the one for $2 \times 2$ grid of lamps, indicating a potential advantage over the architecture of IBM devices. Our experimental results though, did not demonstrate the possibility of solving the computational problem on IQM devices. While the relative frequency for all expected states combined was close to $0.5$ in some cases, we can see that for the expected state \texttt{101} the relative frequency was below $0.125$ in all our experiments.

We checked the runtime usage for our experiments on quantum hardware. On IBM devices, each experiment usually took roughly 3 seconds of runtime, with the exception of Estimator with \reslevel set to 1 and 2 -- a single experiment then took 12 and 15 seconds, respectively. We observed improved results when error mitigation was applied using the Estimator, and we can see an increase in runtime cost. For IQM devices, each of our experiments usually took 1 or 2 seconds. We noticed some exceptions -- sometimes the same circuit (that is actually run on hardware) with the same number of $\texttt{shots}$ can use a noticeably different amount of runtime on quantum hardware. For example, while on a $2 \times 2$ grid with \optlevel 2 used 3 seconds of runtime on \fez~(as well as on the other two IBM devices), there was a case when the experiment used 7 seconds of runtime. A similar case was observed on IQM devices. While the experiment for the SAT instance used up one or two seconds of runtime on all three IQM devices, there was one case, when experiment used 4 seconds on IQM Emerald.

\section{Conclusion}

We designed two circuits that implement Grover's Search with oracles for the Lights Out problem. The instance for the $2 \times 2$ grid of lamps involves 9 qubits, while the instance for M\"obius ladder with 6 lamps involves 16 qubits. While the depth and number of two-qubit operations of the latter circuit do not exceed those of the former, it proved to be a harder challenge for available hardware. The circuit for M\"obius ladder with 6 lamps can serve as an interesting benchmark for upcoming new hardware, as with most optimal settings, we barely managed to surpass 50\% observed success frequency.

The experiments provided a valuable opportunity to observe the progress of available quantum hardware. The current hardware available for free is capable of running circuits that exceed a depth of 700, having more than 300 two-qubit operations. The IQM devices, while solving smaller tasks than IBM devices, are still interesting, as they are capable of solving small instance examples in a practical setting, and are capable of generating more compact circuits for harder problems. For IBM devices, we can see some progress in the year 2023-2024, as our problem instance was solved more reliably and consistently on Heron r2 devices compared to Heron r1 device, and the hardest checked problem becomes almost solvable on Heron r2 device.

\newcommand{\miami}{\texttt{ibm\_miami}}

Mostly, \marrakesh~performed better than the other two IBM devices, while there were often cases when \fez~performed worse than \torino. This highlights that manufacturing quality has a significant impact on QPUs, and newer devices may underperform compared to older ones. 
Calibration has a significant effect on performance, and theoretically ``weakest'' device can surpass the ``strongest'' one. It may be worthwhile to perform an initial experiment on all available devices to verify the calibration outcomes, and then to work on the best-performed one, as our experiments confirmed the consistency of performance difference between devices within the same calibration period. 
Moreover, calibration can lead to unexpectedly good/bad performance, e.g., a new IBM device (that is not available for free) \miami~after one calibration in the evening of January 12, 2026, had a median readout error rate of 33\%, leading to almost random outcomes on any quantum circuit. Usually, it has a readout error rate of approximately 1\%. \fez~had a similar issue on January 23, 2026 (a median readout error rate of 49\%), which was the next day after maintenance.

\section*{Acknowledgment}
The work of M. Dimitrijevs and A. Yakary{\i}lmaz was supported by Latvian Quantum Initiative under European Union Recovery and
Resilience Facility project no. 2.3.1.1.i.0/1/22/I/CFLA/001.

\end{document}